# Virtual Heritage at iGrid 2000

Dave Pape[1], Josephine Anstey[2], Bryan Carter[3],
Jason Leigh[1], Maria Roussou[4], Tim Portlock[1]

[1] University of Illinois at Chicago, USA, [2] University at Buffalo, USA,
[3] Central Missouri State University, USA, [4] Foundation of the Hellenic World, Greece
contact: pape@evl.uic.edu

**Abstract**

As part of the iGrid Research Demonstration at INET 2000, we created two Virtual Cultural Heritage environments – "Virtual Harlem" and "Shared Miletus". The purpose of these applications was to explore possibilities in using the combination of high-speed international networks and virtual reality (VR) displays for cultural heritage education. Our ultimate goal is to enable the construction of tele-immersive museums and classes. In this paper we present an overview of the infrastructure used for these applications, and some details of their construction.

**Introduction**

Cultural heritage is becoming an important application for virtual reality technology. An EC/NSF Advanced Research Workshop identified it as one of the key application domains for driving the development of new human-computer interfaces [1]. Virtual heritage applications use the immersive and interactive qualities of VR to give students or museum visitors access to computer reconstructions of historical sites that would normally be inaccessible, due to location or fragile condition. They also provide the possibility of visiting places that no longer exist at all, or of viewing the how the places would have appeared at different times in history. A number of research projects have delved into the problems of building accurate digital reconstructions of ancient artifacts. For example, Boulanger et al describe the work done in creating models of the tombs of Nefertari and Tutankhamun, and presenting a VR tour of the models in a public museum [2]. Similarly, a recent joint project by the Israel Antiquities Authority and UCLA created an interactive simulation of the Herodian Mount for the Jerusalem Archaeological Park [3].

Tele-immersion is defined as the synthesis of collaborative virtual environments, audio and video conferencing, and supercomputing resources and massive data stores, all interconnected and running over high-speed national or worldwide networks [4]. Tele-immersion enables people at distant locations to work together in a common virtual space, particularly on problems in highly compute-intensive areas such as scientific visualization, computational steering, and design engineering. With tele-immersion technologies, it will be possible to take the cultural heritage work that has been done so far, and expand it beyond a single location to create a virtual museum or educational environment, which can be visited at any time by users on the Internet.

Shared Miletus and Virtual Harlem emerged from very different communities, with different goals and motivations. However the combination of VR and tele-immersion enriches and extends both projects. Both projects document places that no longer exist – the ruins of Miletus are sunk in a swamp near the Turkish coast, the Harlem of the Harlem Rennaissance exists only on celluloid, paper and in music. In both cases a carefully documented virtual environment can bring back a sense of the place, the history, the architecture, and be used as an educational tool.

Shared Miletus is a reconstruction of the ancient city of Miletus, created by the Foundation of the Hellenic World (FHW), a non-profit, privately funded museum and cultural research institution in Athens, Greece. The FHW is an interpretive museum. Its mission is to preserve and present Hellenic history and culture; it seeks to use state-of-the-art technology to accomplish these goals [5].

The Virtual Harlem Project is a collaborative learning network designed to supplement African American literature courses studying the Harlem Renaissance [6]. The project was originally conceived in 1998 by Bryan Carter at Central Missouri State University and the first prototype was initiated in collaboration with Bill Plummer at the Advanced Technology Center at the University of Missouri. In August of 1999, the University of Illinois at Chicago joined the project, to translate the Harlem experience into a fully immersive environment.

**Background**

### CAVE Virtual Reality

Shared Miletus and Harlem were shown on the CAVE[tm] and ImmersaDesk[tm] VR systems at INET 2000. The CAVE (CAVE Automatic Virtual Environment) is a projection-based virtual reality system [7]. In contrast to head-mounted display VR systems, where the user views a virtual world through small video screens attached to a helmet, in projection-based VR large, fixed screens are used to provide a panoramic display without encumbering the user. The CAVE is a 10 foot-cubed room. Stereoscopic images are rear-projected onto the walls creating the illusion that 3D objects exist with the user in the room. The user wears liquid crystal shutter glasses to resolve the stereoscopic imagery. An electromagnetic tracking sensor attached to the glasses allows the CAVE system to determine the location and orientation of the user's head. This information is used by the Silicon Graphics Onyx that drives the CAVE to render the imagery from the user's point of view. The user can physically walk around an object that appears to exist in 3D in the middle of the CAVE. The user holds a wand which is also tracked and has a joystick and three buttons for interaction with the virtual environment. Typically the joystick is used to navigate through environments that are larger than the CAVE itself. The buttons can be used to change modes, or bring up menus in the CAVE, or to grab virtual objects. Speakers are mounted to the top corners of the CAVE structure to provide sounds from the virtual environment. The ImmersaDesk is a smaller VR system with a 6 by 4 foot angled screen, resembling a drafting table, that is also capable of displaying rear-projected stereoscopic images [8]. It has a similar tracking system and the same wand interface. The CAVE and the ImmersaDesk can run the same VR applications.

VR applications displayed on the CAVE and ImmersaDesk systems can be linked over high-speed networks. In these tele-immersive experiences users can share the same virtual world from remote locations. They can interact with each other and with the objects in the virtual world. Users see each other as avatars; simple physical models like puppets, receiving tracking information over the network. For example: a user in CAVE in Boston navigates through the virtual space, turns his, head or points, and a user in a linked CAVE in Chicago sees the avatar move, turn and point. This simple body language makes collaboration in the virtual environment much more effective. The users also speak to each other using audio-confencing tools. Tele-immersion supports high-quality interaction between small groups of participants involved in many fields including design, training, and education.

### CAVERNsoft

CAVERNsoft is a toolkit for building tele-immersive VR applications [9]. Its purpose is to enable rapid generation of tele-immersive applications, without the application authors needing to worry about network protocols and architectures. It is a C++ library that provides a wide range of tools at different levels of complexity. It includes low-level network classes that form interfaces to TCP, UDP, and multicast socket functions, and other classes for threading and cross-platform data conversion. Built on top of these are middle-level modules for such things as remote transfer of very large files, HTTP communications, and remote procedure calls. Above these are database modules that can be used to emulate a distributed shared memory system.

The database module provides a simple two-field database, associating arbitrary chunks of binary data with character string keys. The keys are treated like Unix directory paths, so that a hierarchical arrangement of data is possible. When a client connects to a CAVERN database, it can make asynchronous requests to fetch particular keys' values, and it can store new values for keys. Stored data is automatically reflected to all other clients by the database server. The database client class can also be used without a server, in which case it operates in a standalone mode, making it transparent to the application whether the database is network-shared or not. An additional feature of the database is that data may be stored using either a reliable or an unreliable network connection, under control of the application. This allows one to store state changes, such as a switch being turned on or off, reliably, so that all clients will be sure to receive the change, while storing data that may be a continuous stream, such as avatar positions, unreliably, so that it can be delivered to other clients more quickly.

**iGrid 2000**

The International Grid (iGrid) is a series of research demonstrations highlighting the value of international high-speed computer networks in science, media communications, and education [10]. iGrid 2000 took place at the INET 2000 conference in Yokohama, Japan, and featured twenty-four applications from North America, Europe and Asia displayed on a CAVE, two ImmersaDesks and the Access Grid presentation environment. iGrid was connected to the JGN, the WIDE Project Network (in cooperation with NTT, TTNet and PNJC), APAN and the APAN/TransPAC (100Mbps) link to STARTAP$^{sm}$ in Chicago, Illinois. STARTAP serves as an international connection point for several research networks in America, Europe, and Asia. APAN, the Asia-Pacific Advanced Network is a non-profit international consortium that provides an advanced networking environment for the Asian Pacific research community and promotes international collaboration.

Our virtual heritage applications at iGrid 2000 were networked between the VR displays in Yokohama, a CAVE at the Electronic Visualization Laboratory in Chicago, and the Advanced Technology Center in Missouri.

**Ygdrasil**

Miletus and Harlem are both based on Ygdrasil. Ygdrasil is a framework that we are developing as a tool for creating networked virtual environments. It is focused on building the behaviors of virtual objects from re-usable components, and on sharing the state of an environment through a distributed scene graph mechanism. It is presently being used in the construction of several artistic and educational applications.

Ygdrasil is built in C++, around SGI's IRIS Performer visual simulation toolkit [11] and the CAVERNsoft G2 library. Performer provides a hierarchical representation of the virtual world database, called a scene graph. The scene graph is a tree that encodes the grouping of objects and nesting of 3D transformations, and provides tools for operations such as switching elements on or off. For example, Figure 1 shows a basic scene graph for a robotic puppet, consisting of three geometric models and some transformations.

Ygdrasil focuses on constructing dynamic, interactive virtual worlds, so in addition to the basic graphical data as used in Performer, its scene graph nodes can have behaviors (i.e. functions to update the nodes' state) attached to them. Tying the behaviors to the nodes has made it easy to assemble large worlds from simple components. A behavior is added by taking one of the core classes, such as a transformation node, and deriving a subclass with the new features. Individual node classes are compiled into dynamically loaded objects (DSOs), so that they can be rapidly added to a world or modified. The system includes a number of pre-made classes (also DSOs) that implement common virtual world tools – sounds, users' avatars, navigation controls, triggers that detect when a user enters an area, etc. These built-in tools simplify the quick construction of many basic applications.

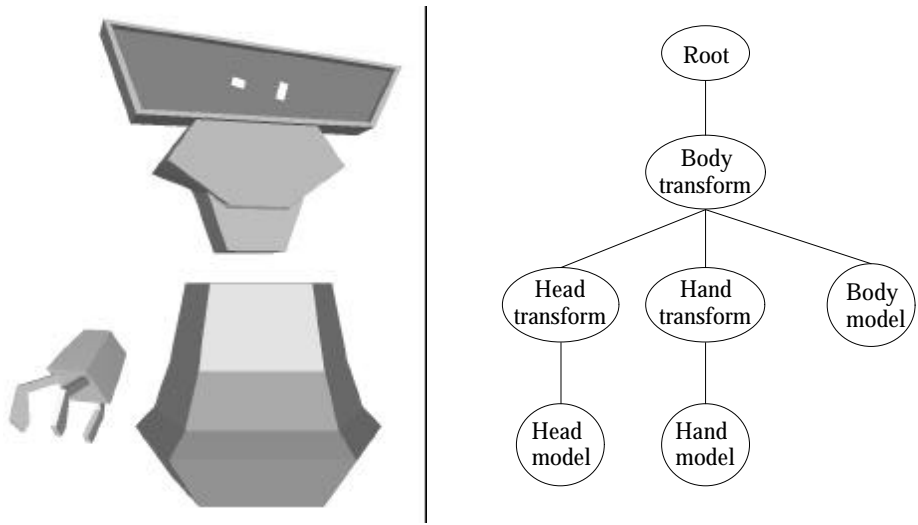

Figure 1. A puppet and its scene graph

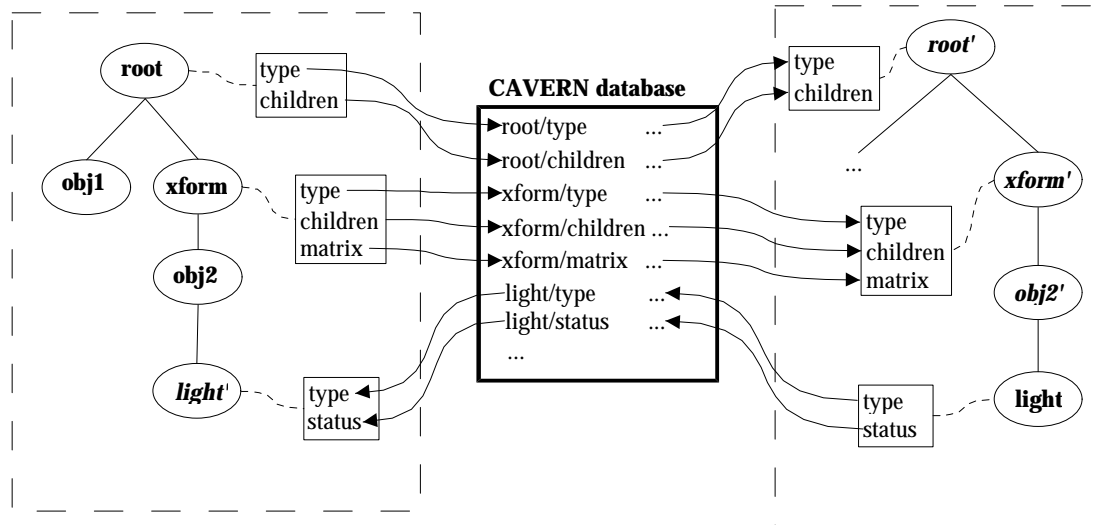

Figure 2. Ygdrasil - Scene graph as a shared database

The Ygdrasil framework further extends the scene graph to be shared across the network. The concept of a shared scene graph has also been explored by other recent VR development systems, such as Avango (formerly known as Avocado) [12]. In a shared system, each participant has his own local copy of the common scene graph. Node data, such as lists of nodes' children, transformation matrices, and model information, are automatically distributed among these participants as the data changes or when participants first join the shared world.

In Ygdrasil, the scene graph is shared using CAVERNsoft's distributed database model. The data that are to be shared for any node in the scene graph are stored in the database keyed by the node name and the data members' names (see Figure 2). The scene can be dynamic, so whenever a client first learns about a new node (by a reference to the node's name), it looks up the node and its type in the database. The client can then create an appropriate local copy of the node, and retrieve all the other data as needed.

Each particular node is considered to be owned by the host that creates it. The owning host maintains the master version of the node, and executes any behavior associated with it. All other hosts will create proxy versions of the node, and only receive data for it through CAVERNsoft; they do not directly modify the node (they can however send messages to the master copy to request changes). The proxy version of a node is typically of the simpler parent class type, without the added behavior code. Thus, if all of the main behaviors for a virtual world are executed by a single master version of the scene, remote sites can join in this world without needing anything beyond the standard, core node types.

**Shared Miletus**

The FHW owns two virtual reality systems, an ImmersaDesk and a ReaCTor (a CAVE-like system), that are used to present a variety of content created by Foundation staff. One of the first VR applications shown at FHW was a reconstruction of Miletus (see Figure 3). Detailed models of some of the buildings of Miletus were created, and museum visitors can explore the city as it was in antiquity. Experienced museum guides lead visitors through the exhibits; the guides have both technical skills to operate the VR displays and museum education skills to explain the history of the city.

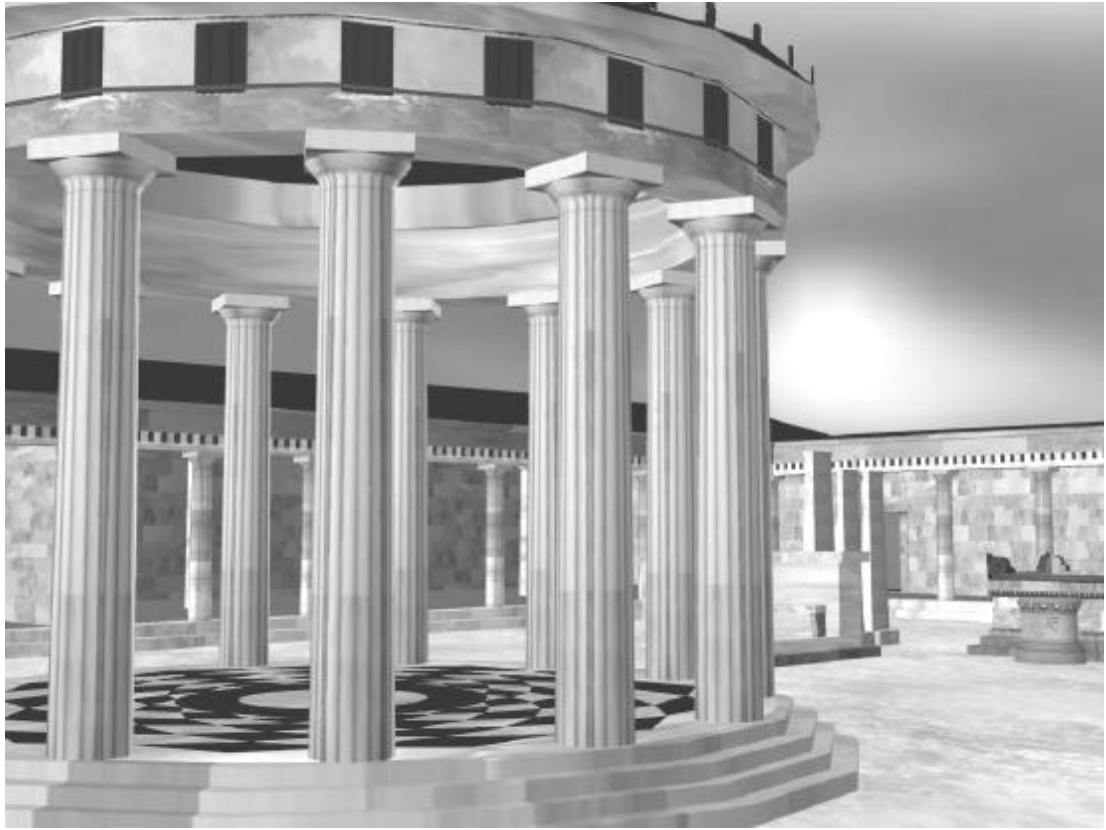

Figure 3. The Delfinio of Miletus

The objective of the Shared Miletus application was to take the content that would normally be shown in the controlled environment of FHW's museum, and let remote, networked people visit it. In particular, we did not want to simply make it something like a VRML model that visitors would download and then play with on their own; instead, it was to be a dynamically shared world, "hosted" by the conference demonstrators.

In creating the demonstration of the Shared Miletus environment, we focused on two issues – guiding visitors through the city, and providing them with information about what they were seeing. These features needed to work in an internationally distributed environment, where users could come and go from the space at will.

Many museum-based VR exhibits lead visitors through the virtual world on a pre-selected path, so that users do not have to learn any special controls or know where they should be going. In our case, we wanted to give the visitors freedom to explore Miletus at their own pace. They were given a 3D wand for simple joystick-driven navigation; a recorded introduction explained how to use the wand. To make it easier to get to places of interest, we also gave them a dynamic, virtual map. This map showed the layout of the city, the user's position in it, and also the positions of any other visitors. This helped them to drive to particular buildings, or to meet up with other visitors or guides from the museum. The map also served as a navigation shortcut – clicking on a particular building summoned a magic carpet that automatically brought the user to the building's entrance.

The first stage of providing visitors with information about Miletus was to include expert human guides. Official guides could enter the shared world, just like ordinary visitors. Through their avatars, and streaming network audio connections, the guides could then interact with the visitors, pointing out special details and answering questions.

Given the international scope of the shared space, we felt that a few human guides alone would not be enough. So, we placed automated information kiosks within the various buildings of Miletus. These kiosks contained recorded audio commentaries describing each building and its history. This audio was available in multiple languages; for the iGrid demo we provided English and Japanese commentaries, but given enough time and translation personnel, any number of languages could be supported. The multi-lingual capability was implemented by having each visitor carry his own virtual audio tool. The tool was effectively a part of the user's avatar, and kept track of his preferred language. When the user approached a kiosk, it detected the presence of an audio tool and sent the tool messages informing it of what recordings the kiosk could provide. If the user chose to listen to one of them, the tool would send a request back to the kiosk, asking for the sound file in the desired language. Other tools at the entrance to the world could be used to switch languages – clicking on a Japanese flag icon would send a message to the user's audio tool to use Japanese, for example. The audio tool also provided the introduction and navigation instructions in the appropriate language.

**Virtual Harlem**

The goal of the Harlem project is to provide an environment that contextualizes the study of the Harlem Renaissance, an important period in African American literary history, through the construction of a virtual reality scenario that represents Harlem, New York, as it existed circa 1925-35. Networked students can navigate through the streets and buildings of Harlem, and see the shops, homes, theaters, churches, and businesses that the writers of the period experienced in their everyday lives. Students can enter the Cotton Club and listen to African American entertainers like Fats Waller or Duke Ellington. They can experience the fact that while the entertainers and servers were African American, the clientele was white. In Virtual Harlem, students can hear the music of the time, watch the dances of the period, and even listen to political speeches of figures like Marcus Garvey or enjoy a poetry reading by Langston Hughes. An instructor at one site can lead students through the environment, explaining things and answering questions, and students can present their own research.

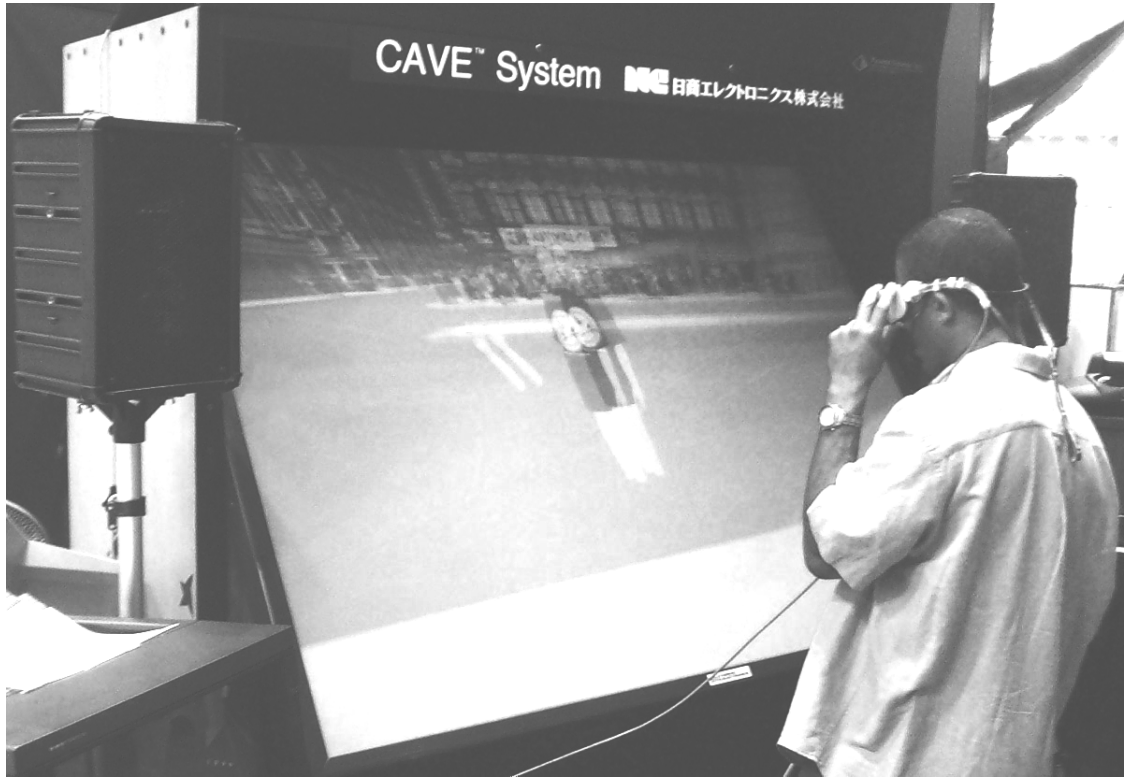

Figure 4. Virtual Harlem on the ImmersaDesk at iGrid

The people working on Virtual Harlem had many discussions about what type of narrative could be developed that could reconcile the presence of early 21$^{st}$ century students and teachers in an early 20$^{th}$ century environment. For the short term we loosely settled on the tour guide narrative. The avatars for Virtual Harlem were made to function as realistic representations of the actual people leading the tours of Harlem replete with early 21$^{st}$ century style clothing.

The particular features of the Harlem environment that were implemented for iGrid 2000 were a period trolley car for transportation, some basic virtual characters, and movies in the Cotton Club. The trolley car moves automatically through the streets. Visitors can board the car by entering it, at which point they are 'attached' to it and move with it; they can exit by stepping outside of the car. Distributed along the city sidewalks are historical characters – Langston Hughes, Marcus Garvey, a group of women headed to a rent party, etc. These characters are billboarded, cutout images of the people from 1920s photographs. When a user approaches the characters, recorded speeches or conversations are played. Visitors can also enter the Cotton Club; inside is a static re-creation of patrons and staff in the main hall, and an interactive movie screen on the stage. By clicking wand buttons, any visitor can play back QuickTime movies of various performances that were at the Cotton Club. The movies are displayed as animated textures attached to a large square that is part of the 3D environment.

**Use of the Network**

In running the shared versions of Miletus and Harlem, the network was used for: distributing data files, sharing the scene graph, and audio conferencing.

A VR environment is typically made up of many individual 3D models (buildings, characters, etc.), texture images for the models, and recorded sound files. The complete Miletus environment comprised roughly 650 files, for a total of 228 megabytes. Harlem was similar – 610 files, totaling 212 megabytes. All of these files needed to be downloaded to the local VR system for any remote user to join the world. The initial draft of Ygdrasil did not include support for distributing the data files automatically, so users would download them in an archive file along with the client program. Newer versions of Ygdrasil can transfer files from web servers if they do not exist locally, allowing a user to join a world little more than the client program. In this case, the user will see the virtual world gradually build up bit by bit, with objects appearing as they are downloaded.

The first thing that a client must do when joining a shared world is to get a copy of the scene graph. This is done by giving it the network address of the machine running the CAVERN database server or reflector. The client requests information about the root node, and then from there builds up a view of the complete scene. Once the program is running, it continues to receive updates for any dynamic objects (e.g. the moving trolley car in Harlem), and send its own updates for the user's avatar, through the shared database. The large part of these environments are static – the amount of data that is changing at any one time is small compared to the size of the entire scene.

Live audio connections are important for distributed users to talk to each other and collaborate. To enable this, we ran an audio-conferencing tool in parallel with the VR programs. This tool is a standard part of the CAVERNsoft library. It is a unicast UDP/IP based program (multicast connections are not available between many of the sites that we work with); a central reflector is used to allow more than two hosts to connect. In addition to its use within the virtual environment, this audio link is often very useful in demo situations such as iGrid, providing a back-channel for setup and problem-solving communications between sites.

**Assessment of Networked Demonstration**

The "Shared Miletus" and "Virtual Harlem" projects ran successfully networked between Yokohama, Japan and North America. Bryan Carter and Maria Roussou acted as expert networked guides and educators in the virtual worlds, discussing the history, architecture and issues of cultural importance that the virtual environments provoked. Roussou also used the pre-recorded Japanese language explanations as she guided tour groups who did not speak English or Greek.

Our most significant problem was the startup time for the applications. It typically took several minutes for client programs to simply create their copy of the shared scene graph, before the user could do anything. This was due to high network latency – the Yokohama to Chicago round trip ping time was 150 milliseconds. The problem is related to the long-fat pipe problem in TCP/IP networking [13]. The LFP problem occurs when the TCP is using a small window, and the system must wait a long time for an acknowledgment for each window of data. In our case, although we were using UDP/IP, there were similar acknowledgment delays during the startup. To build the local scene graph, a proxy for every existing remote node must be created. This requires waiting for a response from the remote system with the node's type information. The creation of nodes was also effectively serialized – until one node's type was received and it's proxy created, no subsequent nodes could be created. With a 150 ms round trip time, at best 6 nodes could be created per second. The Miletus scene graph consists of 1025 nodes, thus leading to the extremely slow startup.

Ygdrasil has since been redesigned to address this problem. Dummy proxies of nodes are created first, without waiting for a response from the shared database. When the type information is actually received, the dummy proxy is transparently replaced by the correct proxy. This allows many key requests to be outstanding at once. Furthermore, as all nodes can have a "children" key, this data is also requested immediately, along with the "type" key. The net result then is that initialization times will be proportional to the depth of the scene graph, rather than the total number of nodes, a significant improvement.

A second difficulty also had to do with high network latencies. Certain activities can involve nodes controlled by separate hosts interacting closely. For example, when a user rides on the trolley car in Harlem, his avatar must constantly update its position based on that of the car, which is being updated by a remote host. Locally, everything will look okay to the user, but other people may see the user's avatar bobbing back and forth within the car, as the position data for the two is received from the different hosts at different times. This problem can be resolved by dynamically manipulating the scene graph; if the user's avatar node is attached below the car node, then everyone will see him moving relative to its coordinate system, and the avatar will not need to constantly update its position just to stay with the car.

**Conclusion and Future Work**

The iGrid 2000 demos of "Shared Miletus" and "Virtual Harlem" were successful; however, the high-end research nature of the venue points out the current limitations of these applications. The general public does not have easy access to high speed networks and large-scale VR devices. We do not feel these problems are insurmountable, though – polygon reduction algorithms and other research can help cut down the size of the data involved in these applications, general broadband network access is continually improving, bring greater bandwidth to private citizens, and home computer capabilities are increasing thanks to the demands of games. We are now able to run reduced versions of Miletus and Harlem on an ordinary Linux PC with a commercial graphics card.

"Shared Miletus" has demonstrated the possible future use of networking and virtual reality to provide a tele-immersive museum. This approach can increase the accessibility of a museum to world-wide audiences. It can also allow much larger and more in-depth exhibits than in a physical museum, and offers a chance for greater direct interaction by visitors with an exhibit. "Virtual Harlem" is an ongoing research project for scholars in the fields of VR and the Harlem Renaissance and is in continuing use as a teaching tool. We have only just scratched the surface of what is possible in such virtual museums and classrooms; much work still remains in both the creation of tools and the construction of new experiences.

One such tool is an annotation system that is currently being deployed in Virtual Harlem. The annotation is based on the "VR Mail" system, originally designed for asynchronous collaboration [14]. This tool allows users to leave 3D messages in the virtual world comprising a voice, gesture and positional recording. In the case of "Virtual Harlem," a lecturer will be able to record a guided tour or a series of notes at one point in time, and students can come to listen to the information and record their own comments at a different time.

**Acknowledgments**

We would like to thank the entire iGrid 2000 staff, especially Hiroshi Esaki and Goro Kunito of the University of Tokyo and Akihiro Tsutsui of NTT, for their help and extreme dedication to making the show a success.

The virtual reality research, collaborations, and outreach programs at the Electronic Visualization Laboratory (EVL) at the University of Illinois at Chicago are made possible by major funding from the

National Science Foundation (NSF), awards EIA-9802090, EIA-9871058, ANI-9980480, and ANI-9730202, as well as the NSF Partnerships for Advanced Computational Infrastructure (PACI) cooperative agreement ACI-9619019 to the National Computational Science Alliance. EVL also receives major funding from the US Department of Energy (DOE) Science Grid program, awards 99ER25388 and 99ER25405, and the DOE ASCI VIEWS program, award B347714. In addition, EVL receives funding from Pacific Interface on behalf of NTT Optical Network Systems Laboratory in Japan.

ImmersaDesk is a registered trademark, and STARTAP is a service mark, of the Board of Trustees of the University of Illinois. ReaCTor is a trademark of Trimension Systems.